\documentclass[aps,prl,showpacs,twocolumn,superscriptaddress]{revtex4}

\usepackage{epsfig}
\usepackage{graphicx}
\usepackage{float}
\usepackage{dcolumn}
\usepackage{amsmath}

\newcommand{\ket}[1]{\left| {#1} \right\rangle}
\newcommand{\bra}[1]{\left\langle {#1}\right |}

\newcommand{\proj}[1]{\left| {#1} \right\rangle \left\langle {#1}\right |}

\begin{document}


\title{Photon polarization entanglement induced by biexciton: 
 experimental evidence for violation of Bell's inequality}

\author{Goro Oohata}
\affiliation{Laboratory for Nanoelectronics and Spintronics, Research Institute of Electrical Communication, 
Tohoku University, Sendai 980-8577, Japan}
\affiliation{ERATO Semiconductor Spintronics Project,
              Japan Science and Technology Agency, Japan}

\author{Ryosuke Shimizu}
\affiliation{CREST, Japan Science and Technology Agency, Japan}

\author{Keiichi Edamatsu}
\affiliation{Laboratory for Nanoelectronics and Spintronics, Research Institute of Electrical Communication, 
Tohoku University, Sendai 980-8577, Japan}
\affiliation{CREST, Japan Science and Technology Agency, Japan}

\begin{abstract}

We have investigated the polarization entanglement between photon pairs
generated from a biexciton in a CuCl single crystal via resonant hyper parametric scattering.
The pulses of a high repetition pump are seen to provide improved
statistical accuracy and the ability to test Bell's inequality.
Our results clearly violate the inequality and thus manifest the quantum entanglement and nonlocality of the photon pairs. 
We also analyzed the quantum state of our photon pairs using quantum state tomography.

\end{abstract}


\maketitle

Recently, the issue of ``quantum entanglement'' has been 
attracting the interest of many researchers, because this property acts
as an essential principle in quantum info-communication
(QIC) technologies. 
The first reliable source that experimentally manifested the entanglement was cascaded two-photon emission from a single atom, such as calcium \cite{Kocher67a, Freedman72a, Aspect81a, Aspect82a} and mercury \cite{Clauser76a,Fry76a}.
In this scheme, the change in the atom's total angular momentum is
transferred to the photon pair, 
so that the photons' polarizations, i.e., internal angular momenta, are entangled.
Aspect {\it et al.}
\cite{Aspect81a,Aspect82a} 
demonstrated the polarization entanglement of photons generated from a calcium atomic cascade
by testing 
Clauser, Horne, Shimony and  Holt (CHSH) type Bell's inequality \cite{Clauser69a}.
The other popular method to generate polarization-entangled photons 
is to use spontaneous parametric down-conversion (SPDC) \cite{Kwiat95a, Kwiat99a}. 
The phase-matching condition concerned with macroscopic coherence of the optical waves
is essential to generate the entanglement in SPDC.
In order to proceed in the development of QIC, semiconductor sources of
entangled photons are highly desired.
Cascaded two-photon emission from a biexciton, 
semiconductor analogue of the atomic cascade,
is a promising method to generate polarization-enetangled photons \cite{Benson00a}. 
Recently, we demonstrated  for the first time entangled photon
generation from a semiconductor material \cite{Edamatsu04a}. 
We used biexciton-resonant hyper parametric scattering (RHPS) 
\cite{Strekalov02a,Savasta99a}
in a CuCl crystal.
The RHPS, or two-photon resonant Raman scattering, in CuCl has been 
thoroughly investigated in view of classical spectroscopy \cite{Itoh78a,Ueta86a}.
Most recently, the generation of entangled photons from semiconductor quantum dots 
has been also reported \cite{Stevenson06a,Young06a,Akopian06a}.

In this letter, we report that highly polarization-entangled photon pairs
can be obtained  with time correlation histograms of enhanced visibility
by using a high repetition rate (1 GHz) pump light system.
Based on the results of polarization correlation measurements, 
we show that Bell's inequality has been clearly violated.
Furthermore, we quantitatively analyze the quantum state of the observed photon pairs 
utilizing quantum state tomography \cite{White99a,James01a}. 

\begin{figure}[b]
    \epsfig{file=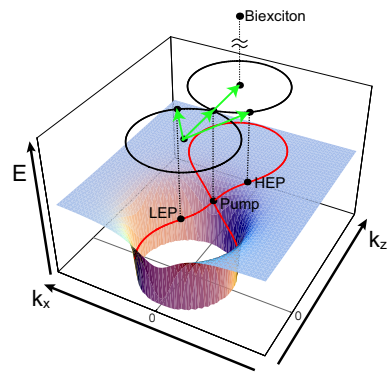, width=50mm }
    \caption{Dispersion relation of exciton polariton (shaded surface) and phase
 matching condition of RHPS (red and black curves: the latter is the projection of the former to the  $k_x$-$k_z$ plane)
 in two-dimensional $k$-space. 
 In our case, the wave vector pump light is
 in parallel with the $k_z$ axis, and the $k_x$-$k_z$ plane corresponds to the
 parallel plane of laboratory system in real space. The green arrows
 indicate the wave vectors of HEP, LEP, and two pump photons.}
    \label{disp}
\end{figure}

\begin{figure}[t]
\begin{center}
   \epsfig{file=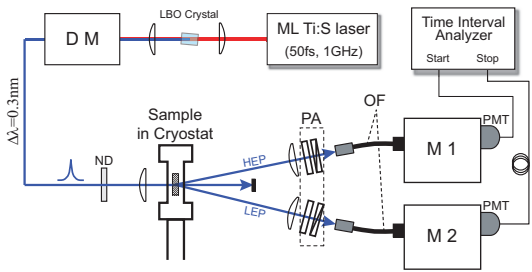, width=70mm}

    \caption{Schematics of our experimental setup.
 DM: zero-dispersion double monochromator. 
 ND: neutral density filter.
 OF: optical multi-mode fiber.  M1, M2: monochromators. 
 PA: polarization analyzers.
 PMT: photon counting photomultiplier.
 }
    \label{setup}
\end{center}
\end{figure}

In RHPS, like in SPDC,  the photons concerned must satisfy the phase-matching condition.
In SPDC, the birefringence of the nonlinear crystal allows photons to
satisfy this condition. 
On the other hand, in RHPS, the dispersion relation
of the exciton polariton is important for the phase-matching condition, 
as shown in Fig. \ref{disp}.
In RHPS, two pump-photons are resonantly absorbed and create a
biexciton in a semiconductor material. 
The created biexciton then decays coherently into two daughter photons.
The difference of RHPS from SPDC is that the process includes a resonant effect,
 making RHPS more efficient than SPDC, 
although RHPS is a higher order ($\chi^{(3)}$) nonlinear process 
than SPDC ($\chi^{(2)}$).
The formation mechanism of polarization entanglement is quite similar to
that of a two-photon cascade emission from a calcium atom \cite{Aspect81a,Aspect82a}.  
That is, the total angular momentum $J$ of the initial state (lowest biexciton)
is $J=0$ and those of the two final states (HEP and LEP) are $J=1$.
Here, HEP and LEP represent a high energy polariton and a low energy
polariton, respectively.
Taking account of the dipole interaction between an exciton and a
photon, the polariton pair are in the entangled angular momentum state
\begin{equation}
  \frac{1}{\sqrt{2}}(\ket{+1, -1} + \ket{-1, +1} ),
\end{equation}
where the first and second symbols in the ket vectors represent the
z-component of $J$ for HEP and LEP, respectively.
Here, we assume that the wave vectors of the two polaritons are
parallel to each other.
Therefore, the emitted photon pair from HEP and LEP are in the
 maximally-entangled polarization state
\begin{equation}
 \begin{split}
  \ket{\Psi} &= \frac{1}{\sqrt{2}}(\ket{R_1 L_2} + \ket{L_1 R_2} ) \\
             &= \frac{1}{\sqrt{2}}(\ket{H_1 H_2} + \ket{V_1 V_2} ) \\
             &= \frac{1}{\sqrt{2}}(\ket{D_1 D_2} + \ket{\bar{D}_1 \bar{D}_2} ),
\label{eq2}
 \end{split}
\end{equation}
where $R_i$ and $L_i$ denote right and left circular polarization,
 $H_i$, $V_i$,  $D_i$ and $\bar{D}_i$ denote  horizontal ($0^\circ$),
 vertical ($90^\circ$),  $45^\circ$ and $-45^\circ$ linear polarization,
respectively.
\begin{figure}[t]
    \epsfig{file=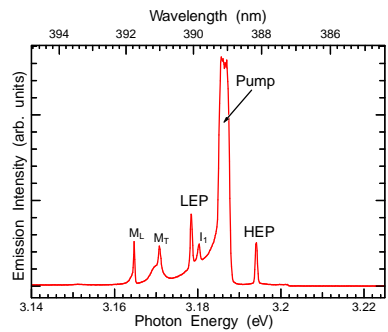, width=50mm }
    \caption{RHPS spectra in CuCl single crystal.
 The central peak indicates the Rayleigh scattered pump light,
 and this energy is tuned at two-photon resonance with the biexciton
 (3.186 eV).
Spectral bandwidth of the pump laser was set to 0.3~nm.
 The two side peaks around the central peak are the RHPS signals of HEP and LEP. 
 The small peak labeled as $I_1$ is the emission of the bound exciton.
 The two peaks ($M_T, M_L$) on the lower energy side are the emission from
 the biexciton leaving the transverse and longitudinal excitons, respectively. 
 }
    \label{spect}
\end{figure}
\begin{figure*}[t]
    \epsfig{file=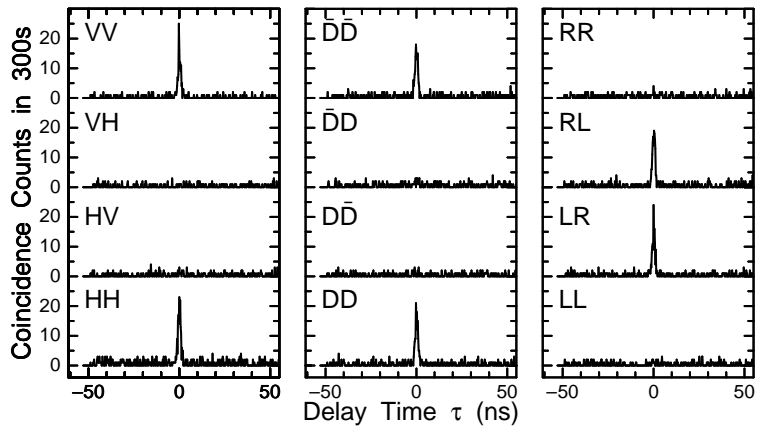, width=100mm }
    \caption{Time correlation histograms between the HEP and LEP for the 
three polarization basis ($H$-$V$, $D$-$\bar{D}$, $R$-$L$).
 The central peak at $\tau = 0$ is the correlated two-photon coincidence
 signal originating from one biexciton.
 The background signal at $\tau \neq  0$ results from the uncorrelated photons.}
    \label{timecorr}
\end{figure*}

The sample used in our experiment is a vapor-phase grown CuCl single crystal having a slab-like
shape (thickness: approx. 100~$\mu$m).
The temperature of the sample was kept at 4 K in a cryostat.
Figure \ref{setup} shows our experimental setup. 
The pump light was the second harmonic light of a 
femtosecond mode-locked Ti:Sapphire laser
having a 1 GHz repetition rate.
The pump light was then spectrally filtered by a
zero-dispersion double monochromator.
The wavelength of the pump light was set to be in the two-photon resonance with the
lowest biexciton. 
The output light from the monochromator passes through an ND filter,
and is focused on the sample.
In order to suppress the accidental coincidence of uncorrelated photons, 
as described below, 
we had to reduce the pump power to around 10~$\mu$W. 
The emitted photons from the sample are fed into the optical multi-mode
fibers connected to the two monochromators. 
The angle between the pump light and the emitted photons was approximately $45^\circ$.
In this condition, the angle between the wave vectors of HEP and LEP inside the crystal was $29^\circ$.
Using the polarization analyzers consisting of a $\lambda/4$ plate and a polarizer
in front of the fiber, 
we measured the polarization state of each photon.
Monochromators (1) and (2) select the HEP and LEP photons
 from the pump laser and the other emissions (see Fig. \ref{spect}).
The photons are detected by two photomultipliers, and the
time-interval analyzer records the difference in arrival time between the
photons.

Figure \ref{timecorr} shows the results of the polarization correlation
measurements using three different polarization 
bases, i.e., $R$-$L$, $H$-$V$, and $D$-$\bar{D}$.
In these results, 
the coincidence signals at $\tau = 0$ clearly appear only in the $RL$, $LR$, $HH$,
$VV$, $DD$, and $\bar{D}\bar{D}$ polarization combinations.
These results indicate that the observed photon-pairs have
polarization-correlation as predicted in Eq. (\ref{eq2}).
It is noteworthy that the signal to noise ratio (S/N) between the coincidence signals
and the uncorrelated background ($\tau\ne 0$)  was quite high;  
in the present study, the S/N was approximately 20.
In contrast,
in the experiment described in our previous report \cite{Edamatsu04a},
 the background had a significant effect on correlated photon
 signals (S/N $\sim$ 2), so that the background was subtracted from the coincidence signal.
In presenting the evidence of the entanglement, this subtraction is an allowable correction.
However, this is not applicable to practical use as an entangled photon source in QIC.
When the pump power is high, an accidental coincidence of
two photons generated from two biexcitons is the main origin of the background. 
In this case,  the accidental coincidence is quadratically proportional to the number of 
signal photons generated per pump pulse.
Thus, the background can be suppressed by the reduction of the pump energy per pulse. 
Thanks to the high repetition rate of the pump laser, 
we were able to suppress the background
while keeping the total number of the signal photons,
as in the data shown in Fig. \ref{timecorr}.
Note that,
in the following,
we analyze the data  
without artificial subtraction of the background.
Using these data with negligibly small background, the violation of Bell's 
inequality 
is demonstrated to show the non-local nature of the state of our
entangled photon pair.
According to CHSH theory \cite{Clauser69a}, the inequality can be written as
\begin{multline}
S = |E(\theta_1,\theta_2) - E(\theta'_1,\theta_2) 
     + E(\theta'_1,\theta'_2) + E(\theta_1,\theta'_2)| \\
 \leq 2,
 \end{multline}
and $E(\theta_1,\theta_2)$ is given by 
\begin{multline}
E(\theta_1,\theta_2) = \\
\frac{C(\theta_1,\theta_2) + C(\theta_1^{\bot},\theta_2^{\bot}) -
C(\theta_1^{\bot},\theta_2)- C(\theta_1,\theta_2^{\bot})}{C(\theta_1,\theta_2) +
C(\theta_1^{\bot},\theta_2^{\bot}) + C(\theta_1^{\bot},\theta_2)+
C(\theta_1,\theta_2^{\bot})},
\end{multline}
where $C(\theta_1,\theta_2)$ is the coincidence count for each
polarization angle and $\theta_i^{\bot}\equiv\theta_i+90^\circ$. 
Table I represents the result of coincidence counts $C(\theta_1,\theta_2)$
recorded for 16 combinations of analyzer setting ($\theta_1$ = $0^\circ$, $45^\circ$,
$90^\circ$, $135^\circ$; $\theta_2$ = $22.5^\circ$, $67.5^\circ$,
$112.5^\circ$, $157.5^\circ$). 
From this result, we can obtain the $S$-value of $2.34 \pm 0.1 > 2$. 
It is clear that the $S$-value apparently violates Bell's inequality by
more than 3 times the standard deviation.

\begin{table}[b]

\begin{ruledtabular}
\begin{tabular}{c|cccc}
 $ \theta_2 \backslash\, \theta_1 $ &$0^\circ$ & $45^\circ$ & $90^\circ$ & $135^\circ$\\
 \hline\\

  $22.5^\circ$ & 134  &  106 &  19  &  44\\
  $67.5^\circ$ & 43 &  107 &  81  & 18\\
  $112.5^\circ$ & 13 &  27  &  85   & 80\\
  $157.5^\circ$ & 104 &  20 &  34  & 143\\
\end{tabular}
\end{ruledtabular}
\caption{\label{bell}Coincidences counts $C(\theta_1,\theta_2)$
for different polarizer angles used in the CHSH inequality test.
They correspond to the integration counts of peak area in 300 seconds.}
\end{table}

\begin{figure}[t]
    \epsfig{file=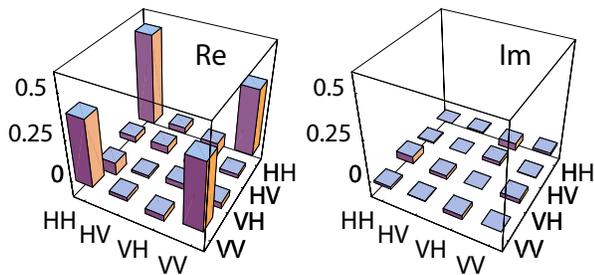, width=78mm }
    \caption{Reconstructed two-photon polarization density matrix for
              the photon pairs generated from RHPS.}
    \label{dens_mat}
\end{figure}

Although the obtained $S$-value violates Bell's inequality, 
the obtained $S$-value was smaller than 
the ideal $S$-value $2\sqrt{2}$ derived from Eq. (\ref{eq2}).
To fully characterize the quantum state of the observed photon pairs,
quantum state tomography was performed to reconstruct the density matrix
of the two-photon polarization state \cite{White99a,James01a}.
For this analysis, 22 independent polarization correlation data including those in
Fig.~4 were used. 
Figure \ref{dens_mat} shows the density matrix thus obtained. 
In this density matrix, 
the two off-diagonal elements,
 $\ket{H_1H_2}\bra{V_1V_2}$ and $\ket{V_1V_2}\bra{H_1H_2}$,
together with the two diagonal elements,
 $\ket{H_1H_2}\bra{H_1H_2}$ and $\ket{V_1V_2}\bra{V_1V_2}$,
 clearly appear, while other elements are almost negligible.
The shape of this density matrix is essentially identical to that
expected from Eq. (\ref{eq2}), in which the two off-diagonal elements represent the
coherence between the two-photon polarization states of $\ket{V_1V_2}$ and
$\ket{H_1H_2}$.
Based on this reconstructed density matrix $\rho$, we estimated the value of
 fidelity, $F \equiv \bra{\Psi} \rho \ket{\Psi}$,
 as 0.85, which is much larger than the classical limit of 0.5.
To quantitatively characterize the degree of disorder and degree of
entanglement of the photon-pair \cite{White02a}, 
we also calculated the linear entropy ($S_L$) and the tangle ($T$) 
from the density matrix.
The value of the entanglement of formation ($EOF$) was also derived from $T$. 
The calculated values are ($S_L, T, EOF$) = (0.31, 0.56, 0.65). 

In the reconstructed density matrix, the element of 
$\ket{H_1 H_2}\bra{H_1 H_2}$ is larger
than that of $\ket{V_1 V_2}\bra{V_1 V_2}$, although they should be
identical based on the ideal maximally entangled state (\ref{eq2}).
 This discrepancy occurs for the following two reasons.
One is the geometrical polarization selection rule in the RHPS process
\cite{Itoh78a, Ueta86a, Comment1}, 
which is unavoidable when the photon pairs are detected at finite angles.
The other is the difference in the transmittance at the sample surface
(between $H$ and $V$ polarized photons) due to Fresnel formula.
Because of the above, the two-photon polarization state of emitted
photon pairs as seen in the sample is written as
\begin{equation}
  \ket{\Psi_\epsilon} = 
  \frac{1}{\sqrt{\epsilon^2+1}}(\epsilon\ket{H_1 H_2} + \ket{V_1 V_2}),
\end{equation}
which is a non-maximally entangled pure state \cite{White99a}. 
In our case, $\epsilon$ is expected to be 1.19.
Furthermore, we should take account of the mixture of the uncorrelated photons.
Thus, the density matrix of the observed two-photon polarization state is
described as 
\begin{equation}
 \rho_{f}= \frac{\gamma}{4}I + (1-\gamma)\proj{\Psi_\epsilon},
\end{equation}
where $\gamma$ is the degree of contribution of the mixed state.
In our case, it is estimated as $\gamma = 0.15$.
With these parameters, 
the value of fidelity, $F \equiv |\sqrt{\sqrt{\rho_f} \rho \sqrt{\rho_f}} |^2$, was estimated as 0.94;
the density matrix $\rho_{f}$ 
reproduces most of the features of the density matrix $\rho$ reconstructed from the observed results.

We have succeeded in obtaining entangled photons via
RHPS by using a pump light with a high repetition rate. 
The high visibility coincidence data 
clearly shows the polarization correlation.
From these data, we reconstructed a density matrix  using quantum state
tomography. 
The density matrix shows that the observed photon pairs are highly
entangled and agree with the results of our theoretical model.
In addition, we demonstrated the violation of Bell's
inequality in regard to the entangled photons generated from a semiconductor.
RHPS using a semiconductor can be employed as a
practical QIC device, acting as an entangled photon source. 
The authors thank M. Hasegawa for his help in preparing samples. 
They are grateful to Professors T. Itoh, H. Ishihara, H. Ohno, H. Kokasa, and Dr. Y. Mitsumori 
for their valuable discussion.
This work was supported in part by 
Strategic Information and Communications R \& D Promotion Program (SCOPE)
of the Ministry of Internal Affairs and and Communications,
and by 
a Grant-in-Aid for Creative Scientific Research (17GS1204) 
of the Japan Society for the Promotion of Science.


\begin{thebibliography}{22}
\expandafter\ifx\csname natexlab\endcsname\relax\def\natexlab#1{#1}\fi
\expandafter\ifx\csname bibnamefont\endcsname\relax
  \def\bibnamefont#1{#1}\fi
\expandafter\ifx\csname bibfnamefont\endcsname\relax
  \def\bibfnamefont#1{#1}\fi
\expandafter\ifx\csname citenamefont\endcsname\relax
  \def\citenamefont#1{#1}\fi
\expandafter\ifx\csname url\endcsname\relax
  \def\url#1{\texttt{#1}}\fi
\expandafter\ifx\csname urlprefix\endcsname\relax\def\urlprefix{URL }\fi
\providecommand{\bibinfo}[2]{#2}
\providecommand{\eprint}[2][]{\url{#2}}

\bibitem[{\citenamefont{Kocher and Commins}(1967)}]{Kocher67a}
\bibinfo{author}{\bibfnamefont{C.~A.} \bibnamefont{Kocher}} \bibnamefont{and}
  \bibinfo{author}{\bibfnamefont{E.~D.} \bibnamefont{Commins}},
  \bibinfo{journal}{Phys. Rev. Lett.} \textbf{\bibinfo{volume}{18}},
  \bibinfo{pages}{575} (\bibinfo{year}{1967}).

\bibitem[{\citenamefont{Freedman and Clauser}(1972)}]{Freedman72a}
\bibinfo{author}{\bibfnamefont{S.~J.} \bibnamefont{Freedman}} \bibnamefont{and}
  \bibinfo{author}{\bibfnamefont{J.~F.} \bibnamefont{Clauser}},
  \bibinfo{journal}{Phys. Rev. Lett.} \textbf{\bibinfo{volume}{28}},
  \bibinfo{pages}{938} (\bibinfo{year}{1972}).

\bibitem[{\citenamefont{Aspect et~al.}(1981)\citenamefont{Aspect, Grangier, and
  Roger}}]{Aspect81a}
\bibinfo{author}{\bibfnamefont{A.}~\bibnamefont{Aspect}},
  \bibinfo{author}{\bibfnamefont{P.}~\bibnamefont{Grangier}}, \bibnamefont{and}
  \bibinfo{author}{\bibfnamefont{G.}~\bibnamefont{Roger}},
  \bibinfo{journal}{Phys. Rev. Lett.} \textbf{\bibinfo{volume}{47}},
  \bibinfo{pages}{460} (\bibinfo{year}{1981}).

\bibitem[{\citenamefont{Aspect et~al.}(1982)\citenamefont{Aspect, Grangier, and
  Roger}}]{Aspect82a}
\bibinfo{author}{\bibfnamefont{A.}~\bibnamefont{Aspect}},
  \bibinfo{author}{\bibfnamefont{P.}~\bibnamefont{Grangier}}, \bibnamefont{and}
  \bibinfo{author}{\bibfnamefont{G.}~\bibnamefont{Roger}},
  \bibinfo{journal}{Phys. Rev. Lett.} \textbf{\bibinfo{volume}{49}},
  \bibinfo{pages}{91} (\bibinfo{year}{1982}).

\bibitem[{\citenamefont{Clauser}(1976)}]{Clauser76a}
\bibinfo{author}{\bibfnamefont{J.~F.} \bibnamefont{Clauser}},
  \bibinfo{journal}{Phys. Rev. Lett.} \textbf{\bibinfo{volume}{36}},
  \bibinfo{pages}{1223} (\bibinfo{year}{1976}).

\bibitem[{\citenamefont{Fry and Thompson}(1976)}]{Fry76a}
\bibinfo{author}{\bibfnamefont{E.~S.} \bibnamefont{Fry}} \bibnamefont{and}
  \bibinfo{author}{\bibfnamefont{R.~C.} \bibnamefont{Thompson}},
  \bibinfo{journal}{Phys. Rev. Lett.} \textbf{\bibinfo{volume}{37}},
  \bibinfo{pages}{465} (\bibinfo{year}{1976}).

\bibitem[{\citenamefont{Clauser et~al.}(1969)\citenamefont{Clauser, Horne,
  Shimony, and Holt}}]{Clauser69a}
\bibinfo{author}{\bibfnamefont{J.~F.} \bibnamefont{Clauser}},
  \bibinfo{author}{\bibfnamefont{M.~A.} \bibnamefont{Horne}},
  \bibinfo{author}{\bibfnamefont{A.}~\bibnamefont{Shimony}}, \bibnamefont{and}
  \bibinfo{author}{\bibfnamefont{R.~A.} \bibnamefont{Holt}},
  \bibinfo{journal}{Phys. Rev. Lett.} \textbf{\bibinfo{volume}{23}},
  \bibinfo{pages}{880} (\bibinfo{year}{1969}).

\bibitem[{\citenamefont{Kwiat et~al.}(1995{\natexlab{a}})\citenamefont{Kwiat,
  Mattle, Weinfurter, Zeilinger, Sergienko, and Shih}}]{Kwiat95a}
\bibinfo{author}{\bibfnamefont{P.~G.} \bibnamefont{Kwiat}},
  \bibinfo{author}{\bibfnamefont{K.}~\bibnamefont{Mattle}},
  \bibinfo{author}{\bibfnamefont{H.}~\bibnamefont{Weinfurter}},
  \bibinfo{author}{\bibfnamefont{A.}~\bibnamefont{Zeilinger}},
  \bibinfo{author}{\bibfnamefont{A.~V.} \bibnamefont{Sergienko}},
  \bibnamefont{and} \bibinfo{author}{\bibfnamefont{Y.}~\bibnamefont{Shih}},
  \bibinfo{journal}{Phys. Rev. Lett.} \textbf{\bibinfo{volume}{75}},
  \bibinfo{pages}{4337} (\bibinfo{year}{1995}{\natexlab{a}}).

\bibitem[{\citenamefont{Kwiat et~al.}(1995{\natexlab{b}})\citenamefont{Kwiat,
  Waks, White, Appelbaum, and Eberhard}}]{Kwiat99a}
\bibinfo{author}{\bibfnamefont{P.~G.} \bibnamefont{Kwiat}},
  \bibinfo{author}{\bibfnamefont{E.}~\bibnamefont{Waks}},
  \bibinfo{author}{\bibfnamefont{A.~G.} \bibnamefont{White}},
  \bibinfo{author}{\bibfnamefont{I.}~\bibnamefont{Appelbaum}},
  \bibnamefont{and} \bibinfo{author}{\bibfnamefont{P.~H.}
  \bibnamefont{Eberhard}}, \bibinfo{journal}{Phys. Rev. A}
  \textbf{\bibinfo{volume}{60}}, \bibinfo{pages}{R773}
  (\bibinfo{year}{1995}{\natexlab{b}}).

\bibitem[{\citenamefont{Benson et~al.}(2000)\citenamefont{Benson, Santori,
  Pelton, and Yamamoto}}]{Benson00a}
\bibinfo{author}{\bibfnamefont{O.}~\bibnamefont{Benson}},
  \bibinfo{author}{\bibfnamefont{C.}~\bibnamefont{Santori}},
  \bibinfo{author}{\bibfnamefont{M.}~\bibnamefont{Pelton}}, \bibnamefont{and}
  \bibinfo{author}{\bibfnamefont{Y.}~\bibnamefont{Yamamoto}},
  \bibinfo{journal}{Phys. Rev. Lett.} \textbf{\bibinfo{volume}{84}},
  \bibinfo{pages}{2513} (\bibinfo{year}{2000}).

\bibitem[{\citenamefont{Edamatsu et~al.}(2004)\citenamefont{Edamatsu, Oohata,
  Shimizu, and Itoh}}]{Edamatsu04a}
\bibinfo{author}{\bibfnamefont{K.}~\bibnamefont{Edamatsu}},
  \bibinfo{author}{\bibfnamefont{G.}~\bibnamefont{Oohata}},
  \bibinfo{author}{\bibfnamefont{R.}~\bibnamefont{Shimizu}}, \bibnamefont{and}
  \bibinfo{author}{\bibfnamefont{T.}~\bibnamefont{Itoh}},
  \bibinfo{journal}{Nature} \textbf{\bibinfo{volume}{431}},
  \bibinfo{pages}{167} (\bibinfo{year}{2004}).

\bibitem[{\citenamefont{Strekalov and Dowling}(2002)}]{Strekalov02a}
\bibinfo{author}{\bibfnamefont{D.}~\bibnamefont{Strekalov}} \bibnamefont{and}
  \bibinfo{author}{\bibfnamefont{J.}~\bibnamefont{Dowling}},
  \bibinfo{journal}{J. Mod. Opt.} \textbf{\bibinfo{volume}{49}},
  \bibinfo{pages}{519} (\bibinfo{year}{2002}).

\bibitem[{\citenamefont{Savasta et~al.}(1999)\citenamefont{Savasta, Martino,
  and Girlanda}}]{Savasta99a}
\bibinfo{author}{\bibfnamefont{S.}~\bibnamefont{Savasta}},
  \bibinfo{author}{\bibfnamefont{G.}~\bibnamefont{Martino}}, \bibnamefont{and}
  \bibinfo{author}{\bibfnamefont{R.}~\bibnamefont{Girlanda}},
  \bibinfo{journal}{Solid State Commun.} \textbf{\bibinfo{volume}{111}},
  \bibinfo{pages}{495} (\bibinfo{year}{1999}).

\bibitem[{\citenamefont{Itoh and Suzuki}(1978)}]{Itoh78a}
\bibinfo{author}{\bibfnamefont{T.}~\bibnamefont{Itoh}} \bibnamefont{and}
  \bibinfo{author}{\bibfnamefont{T.}~\bibnamefont{Suzuki}},
  \bibinfo{journal}{J. Phys. Soc. Japan} \textbf{\bibinfo{volume}{45}},
  \bibinfo{pages}{1939} (\bibinfo{year}{1978}).

\bibitem[{\citenamefont{Ueta and {\it et al}.}(1986)}]{Ueta86a}
\bibinfo{author}{\bibfnamefont{M.}~\bibnamefont{Ueta}} \bibnamefont{and}
  \bibinfo{author}{\bibnamefont{{\it et al}.}}, \emph{\bibinfo{title}{Excitonic
  Processes in Solids}} (\bibinfo{publisher}{Springer, Berlin},
  \bibinfo{year}{1986}).

\bibitem[{\citenamefont{Stevenson et~al.}(2006)\citenamefont{Stevenson, Young,
  Atkinson, Cooper, Ritchie, and Shields}}]{Stevenson06a}
\bibinfo{author}{\bibfnamefont{R.}~\bibnamefont{Stevenson}},
  \bibinfo{author}{\bibfnamefont{R.}~\bibnamefont{Young}},
  \bibinfo{author}{\bibfnamefont{P.}~\bibnamefont{Atkinson}},
  \bibinfo{author}{\bibfnamefont{K.}~\bibnamefont{Cooper}},
  \bibinfo{author}{\bibfnamefont{D.}~\bibnamefont{Ritchie}}, \bibnamefont{and}
  \bibinfo{author}{\bibfnamefont{A.}~\bibnamefont{Shields}},
  \bibinfo{journal}{Nature} \textbf{\bibinfo{volume}{439}},
  \bibinfo{pages}{179} (\bibinfo{year}{2006}).

\bibitem[{\citenamefont{Young et~al.}(2006)\citenamefont{Young, Stevenson,
  Atkinson, Cooper, Ritchie, and Shields}}]{Young06a}
\bibinfo{author}{\bibfnamefont{R.}~\bibnamefont{Young}},
  \bibinfo{author}{\bibfnamefont{R.}~\bibnamefont{Stevenson}},
  \bibinfo{author}{\bibfnamefont{P.}~\bibnamefont{Atkinson}},
  \bibinfo{author}{\bibfnamefont{K.}~\bibnamefont{Cooper}},
  \bibinfo{author}{\bibfnamefont{D.}~\bibnamefont{Ritchie}}, \bibnamefont{and}
  \bibinfo{author}{\bibfnamefont{A.}~\bibnamefont{Shields}},
  \bibinfo{journal}{New J. of Phys.} \textbf{\bibinfo{volume}{8}},
  \bibinfo{pages}{29} (\bibinfo{year}{2006}).

\bibitem[{\citenamefont{Akopian et~al.}(2006)\citenamefont{Akopian, Lindner,
  Poem, Berlatzky, Avron, Gershoni, Gerardot, and Petroff}}]{Akopian06a}
\bibinfo{author}{\bibfnamefont{N.}~\bibnamefont{Akopian}},
  \bibinfo{author}{\bibfnamefont{N.}~\bibnamefont{Lindner}},
  \bibinfo{author}{\bibfnamefont{E.}~\bibnamefont{Poem}},
  \bibinfo{author}{\bibfnamefont{Y.}~\bibnamefont{Berlatzky}},
  \bibinfo{author}{\bibfnamefont{J.}~\bibnamefont{Avron}},
  \bibinfo{author}{\bibfnamefont{D.}~\bibnamefont{Gershoni}},
  \bibinfo{author}{\bibfnamefont{B.}~\bibnamefont{Gerardot}}, \bibnamefont{and}
  \bibinfo{author}{\bibfnamefont{P.}~\bibnamefont{Petroff}},
  \bibinfo{journal}{Phys. Rev. Lett.} \textbf{\bibinfo{volume}{96}},
  \bibinfo{pages}{130501} (\bibinfo{year}{2006}).

\bibitem[{\citenamefont{White et~al.}(1999)\citenamefont{White, James,
  Eberhard, and Kwiat}}]{White99a}
\bibinfo{author}{\bibfnamefont{A.~G.} \bibnamefont{White}},
  \bibinfo{author}{\bibfnamefont{D.~F.~V.} \bibnamefont{James}},
  \bibinfo{author}{\bibfnamefont{P.~H.} \bibnamefont{Eberhard}},
  \bibnamefont{and} \bibinfo{author}{\bibfnamefont{P.~G.} \bibnamefont{Kwiat}},
  \bibinfo{journal}{Phys. Rev. Lett.} \textbf{\bibinfo{volume}{83}},
  \bibinfo{pages}{3103} (\bibinfo{year}{1999}).

\bibitem[{\citenamefont{James et~al.}(2001)\citenamefont{James, Kwiat, Munro,
  and White}}]{James01a}
\bibinfo{author}{\bibfnamefont{D.~F.~V.} \bibnamefont{James}},
  \bibinfo{author}{\bibfnamefont{P.~G.} \bibnamefont{Kwiat}},
  \bibinfo{author}{\bibfnamefont{W.~J.} \bibnamefont{Munro}}, \bibnamefont{and}
  \bibinfo{author}{\bibfnamefont{A.~G.} \bibnamefont{White}},
  \bibinfo{journal}{Phys. Rev. A} \textbf{\bibinfo{volume}{64}},
  \bibinfo{pages}{052312} (\bibinfo{year}{2001}).

\bibitem[{\citenamefont{White et~al.}(2002)\citenamefont{White, James, Munro,
  and Kwiat}}]{White02a}
\bibinfo{author}{\bibfnamefont{A.~G.} \bibnamefont{White}},
  \bibinfo{author}{\bibfnamefont{D.~F.~V.} \bibnamefont{James}},
  \bibinfo{author}{\bibfnamefont{W.~J.} \bibnamefont{Munro}}, \bibnamefont{and}
  \bibinfo{author}{\bibfnamefont{P.~G.} \bibnamefont{Kwiat}},
  \bibinfo{journal}{Phys. Rev. A} \textbf{\bibinfo{volume}{65}},
  \bibinfo{pages}{012301} (\bibinfo{year}{2002}).

\bibitem[{Com()}]{Comment1}
\bibinfo{note}{Degree of polarization of the RHPS signal is given by $
  D(\beta)\equiv(I_V-I_H)/(I_V+I_H)=\left(\frac{2}{\sin^2\beta}-1 \right)^{-1},
  $ where $I_V$ and $I_H$ are the count of $V$ and $H$ polarized photons of HEP
  or LEP, respectively, $\beta$ is the angle between the wave vector of HEP and
  LEP in the sample. Therefore, the two photon polarization state via RHPS
  becomes $ \ket{\Psi(\beta)} = (\sqrt{1-D(\beta)}\ket{H_1, H_2} +
  \sqrt{1+D(\beta)}\ket{V_1, V_2})/\sqrt{2}. $ In our case, $\beta \approx
  29^\circ$, $D \approx 1.4$.}

\end{thebibliography}
\end{document}